\documentclass[twocolumn,superscriptaddress,amsmath,amssymb,aps,prb,longbibliography]{revtex4-2}
\usepackage{graphicx,bm,natbib}
\usepackage{amssymb}
\usepackage{amsmath}
\UseRawInputEncoding
\usepackage{color}
\usepackage{slashed}
\usepackage{float}
\usepackage{tikz}
\usepackage{pgf}
\usepackage{pgfplots}
\usepackage[hidelinks]{hyperref}
\hypersetup{
    colorlinks=false,
    }

\renewcommand{\d}{\mathrm{d}}

\usepackage{amsmath,amsfonts,amssymb}
\newcommand{\be}{\begin{equation}}
	\newcommand{\ee}{\end{equation}}
\newcommand{\bea}{\begin{eqnarray}}
	\newcommand{\eea}{\end{eqnarray}}
\newcommand{\ba}{\begin{array}}
	\newcommand{\ea}{\end{array}}
\newcommand{\cb}[1]{\textcolor{black}{#1}}

\makeatletter
\let\Hy@backout\@gobble
\makeatother

\begin{document}

	\title{Geometric phase for nonlinear oscillators from perturbative renormalization group}

	\author{D. A. Khromov}
        \affiliation{Moscow Institute of Physics and Technology, Dolgoprudny, Moscow Region, 141701, Russia}
	\author{M. S. Kryvoruchko}
        \affiliation{Leipzig University, Leipzig, 04109, Germany }
	\author{D. A. Pesin}
	\affiliation{Department of Physics, University of Virginia, Charlottesville, Virginia, 22904, USA}

	\date{\today}
	
	\begin{abstract}
		We formulate a renormalization group approach to a general nonlinear oscillator problem. The approach is based on the exact group law obeyed by solutions of the corresponding ordinary differential equation. We consider both the autonomous models with time-independent parameters, as well as nonautonomous models with slowly varying parameters. We show that the renormalization group equations for the nonautonomous case can be used to determine the geometric phase acquired by the oscillator during the change of its parameters. We illustrate the obtained results by applying them to the Van der  Pol, and Van der Pol-Duffing models. 
	
	\end{abstract}

	\maketitle

	\section{Introduction} 

It was shown in \cb{Refs.~\cite{Kepler1991phase,kagan1991phase,Haken1992berryphase}} that classical dissipative systems with a limit cycle admit the notion of a geometric phase accumulated when the system's parameters undergo a slow change. This geometric phase is analogous to the Hannay angle~\cite{hannay1985angle} in classical Hamiltonian systems with adiabatic invariants, and to the Berry phase~\cite{berry1984phase} in quantum mechanics. In this work, we show how to calculate this phase for models describable with nonlinear oscillators based on the renormalization group (RG) approach. 

In the case of classical and quantum Hamiltonian systems, the adiabatic theorem~\cite{shankar2012qm} dictates that the system remains on an invariant torus during the slow evolution of its parameters, and either the Hamiltonian equations for the action-angle variables~\cite{LL1} in the classical case or the Schr{\"o}dinger equation in the quantum case determine the dynamics on this torus. Then it can be shown that geometric contributions to appropriately defined phase shifts  appear~\cite{shapere1989phases}. These shifts are independent of how exactly the change in systems parameters is parametrized with time, and only depend on the trajectory that the system executes in the parameter space, hence the name ``geometric". 

The description of the geometric phase for classical dissipative systems with a limit cycle follows the same route as its counterpart for Hamiltonian systems. The very existence of a stable limit cycle replaces the adiabatic theorem for the Hamiltonian systems, and if one can find a way to describe the dynamics of the limit cycle, they can also obtain the geometric phase~\cite{kepler1992review}. In the case of classical dissipative systems, there have been studies based on either numerical experiments~\cite{kepler1991geometric} or their Hamiltonianization~\cite{chakraborty2018hannay}, with subsequent switching to appropriately defined action-angle variables. 

In this work, we show that the renormalization group theory for nonlinear oscillators provides a natural way to transition from the underlying differential equations to the ``coarse-grained" description of slow amplitude evolution along a limit cycle, and extract the geometric phase. 

The possibility to apply the renormalization group as discovered in the quantum field theory, in particular, its formulation due to Bogoliubov and Shirkov~\cite{bogoliubovshirkovbook}, to problems in mathematical physics was noted in Ref.~\cite{Mnatsakanyan1982rg}, and independently in Ref.~\cite{goldenfeld1989rg}. The renormalization group method turned out to be a powerful way to regularize secular terms in the perturbation theory for dynamical systems, and study their asymptotic behavior, see Refs.~\cite{shirkov1988rg,goldenfeldbook} for pedagogical discussions, and historical context. As demonstrated in Ref.~\cite{goldenfeld1996rg}, the method is extremely versatile, and can in principle replace most known methods of asymptotic analysis, such as multiple-scale analysis, time-averaging techniques, and so on. Several examples of the application of the method are given in Refs.~\cite{kirkinis2010,kirkinis2012rg}. The geometric meaning of the renormalization group procedure as a determination of the envelope of perturbative solutions was given by Kunihiro in Ref.~\cite{kunihiro1995envelope}. The mathematical foundations of the method based on the invariant manifold theory, as well as its relation to Wilsonian-type RG, were expounded in Ref.~\cite{EFK}.
    
While the ideas expounded in early works on the subject are invaluable, the ways in which they were applied to specific problems sometimes appear \textit{ad hoc}. In particular, while dealing with the RG equations to linear order in a small parameter is straightforward in various approaches, higher-order corrections are extremely cumbersome, and are obtained in non-systematic ways.  This issue was addressed in Ref.~\cite{Oono} via the so-called ``proto-RG," in which the RG equations are made to be second order in time derivatives, but can then be easily iterated order by order. Previously, similar goals were accomplished in Ref.~\cite{goto1999lie} using a Lie-group based approach, which appears similar in spirit to the one of this work.

In this work, we present a formulation of the renormalization group procedure for nonlinear oscillators based on the exact group law obeyed by the solutions of the corresponding differential equations, similar to the approach of \cb{Ref.~\cite{Mnatsakanyan1982rg,EFK}}. As a result, we are able to show that the nonlinear oscillator models are perturbatively renormalizable to any order. The renormalization group equations are obtained directly from the linear in time secular terms at the prime frequency of the unperturbed oscillator in a completely mechanistic way. We then use these equations to calculate the geometric phase shifts for nonlinear oscillators. 

The rest of the paper is organized as follows. In Section~\ref{sec:RG} we present a general formulation of the renormalization group treatment of nonlinear oscillators. In Section~\ref{sec:phase} we show how to determine the geometric phase for classical dissipative systems from the renormalization group equations. In Section~\ref{sec:VdPoscillators} we apply the developed formalism to two well-known models: Van der Pol, and Van der Pol-Duffing oscillators, and calculate the geometric phase accumulated during a slow change in the limit cycle. We summarize our results in Section~\ref{sec:discussion}. 
	
\section{Perturbative renormalization group for nonlinear oscillators }\label{sec:RG}
    In this work, we are going to study the asymptotic behavior of a nonlinear oscillator, defined via the following equation of motion:
    \begin{equation}\label{eq:generalode}
		\ddot{y} + \omega^2 y = \epsilon (t) f(y,\dot{y}).
	\end{equation}
	In the above equation, a dot over a function implies differentiation with respect to time $t$, and $f(y,\dot{y})$ is an analytic function of both of its arguments.
    For simplicity, at this point we have introduced a single time-dependent small parameter $\epsilon(t)$.  In actual applications we will consider systems with at least two small parameters, such that the system can be driven though a closed cycle on nonzero area in the parameter space, and possibly accumulate a geometric phase in this process. The frequency of the oscillator can also be made time dependent, and serve as a parameter to generate a geometric phase. 
    
    It is well known that the naive perturbation theory for model~\eqref{eq:generalode} developed in powers of $\epsilon$ is in general singular because of the appearance of secular terms, whose magnitude grows with time. This invalidates the perturbation theory at sufficiently long times. We will resort to the renormalization group approach to improve the perturbation theory.

	 \cb{Below we present a formulation of the perturbative RG for nonlinear oscillators, which stems from a group law obeyed by the exact solutions of the corresponding differential equation. Similar general treatments have already been implemented in the literature for the general problem of renormalization-group reduction~\cite{EFK,Oono}. In this work we focus on giving a detailed account of how to apply the method to nonlinear oscillators, suggesting a way to construct perturbative solutions in such a way as to obtain the corresponding RG equations by a completely mechanistic application of the perturbation theory to the needed order. The procedure appears to be even simpler than the proto-RG of Ref.~\cite{Oono}. We did check that for the specific problems considered in this work the two approaches yield the same results. Furthermore, we generalize the RG approach to treat situations with time-dependent coefficients in non-linearities to study geometric phases. }

	\subsection{Adiabatic RG equations}\label{sec:adiabaticRG}
	
	First, we formally neglect the time-dependence of $\epsilon$ in Eq.~\eqref{eq:generalode}. This way we obtain the RG equations that contain only an instantaneous value of $\epsilon$. In this sense, we are going to perform the RG analysis of the following model: 
	\begin{equation}\label{eq:autonomousmodel}
		\ddot{y} + \omega^2 y = \epsilon f(y,\dot{y}),
	\end{equation}
	We will refer to the corresponding RG equations as adiabatic. Nonadiabatic corrections to them, which are linear in $\dot{\epsilon}$, are obtained in Section~\ref{sec:nonadiabaticRG}. 
	
	Let us introduce a notation for a solution of equation~Eq.~\eqref{eq:autonomousmodel}, $y(t,t_0;y(t_0),\dot{y}(t_0))$, which specifies the initial time $t_0$, the observation time $t$, as well as the initial conditions, $y(t_0)$, $\dot{y}(t_0)$. For brevity, we will also use the notation $y(t)$ for this solution, in which the initial time and the Cauchy data are suppressed, but implied. In particular, $y(t_0,t_0;y(t_0),\dot{y}(t_0))=y(t_0)$. 
	
	If a unique solution exists, it must satisfy the following group law for $t>t_1>t_0$:
	\begin{align}\label{eq:grouplaw}
	    y(t,t_1;y(t_1),\dot{y}(t_1))=y(t,t_0;y(t_0),\dot{y}(t_0)),
	\end{align}
\cb{which states that if the values of $y(t_1,t_0;y(t_0),\dot{y}(t_0))\equiv y(t_1)$ and $\dot{y}(t_1,t_0;y(t_0),\dot{y}(t_0))\equiv \dot{y}(t_1)$ are used as the initial conditions for evolution starting at $t_1$, then for $t>t_1$ the system will follow the same trajectory as $y(t,t_0;y(t_0),\dot{y}(t_0))$.} The group law~\eqref{eq:grouplaw} can be used to improve the perturbation theory much the same way it is done in the conventional RG schemes in field theory~\cite{shirkov1984rg,delamotte2004rg}.
	
	It is easy to see that the very existence of the group law implies that 
	\begin{align}\label{eq:globalderivative}
	    \frac{dy(t,t_1;y(t_1),\dot{y}(t_1))}{dt_1}=0.
	\end{align}
	The validity of Eq.~\eqref{eq:globalderivative} is apparent from Eq.~\eqref{eq:grouplaw}, in which $t_1$ does not appear on the right hand side. Equation~\eqref{eq:globalderivative}, valid for general $t_1$, in particular holds for $t_1\to t$:
	\begin{align}\label{eq:localderivative}
	    \frac{dy(t,t_1;y(t_1),\dot{y}(t_1))}{dt_1}|_{t_1\to t}=0.
	\end{align}
 
	 The crucial observation is that the validity of Eq.~\eqref{eq:localderivative}, \emph{enforced for all times $t$}, is sufficient for both Eq.~\eqref{eq:globalderivative} and the group law~\eqref{eq:grouplaw} to hold. Indeed, Eq.~\eqref{eq:localderivative} follows from the following Taylor expansion of $y(t,t_1;y(t_1),\dot{y}(t_1))$ near $t_1=t$:
	\begin{align}
	    y(t,t_1;y(t_1),\dot{y}(t_1))= y(t_1)+\dot{y}(t_1)(t-t_1)+O((t-t_1)^2). 
	\end{align}
 
    The derivative of the right-hand side with respect to $t_1$ vanishes for $t\to t_1$ for any $t$ if, and only if, the initial conditions for $y(t,t_1;y(t_1),\dot{y}(t_1))$ are chosen along an actual solution for $y$. Then it is clear that Eq.~\eqref{eq:localderivative} is equivalent to the group law~\eqref{eq:grouplaw}, since finite evolution along an actual solution of the ODE can be accomplished through a number of infinitesimal steps, for each of which Eq.~\eqref{eq:localderivative} ensures that one moves along the actual solution. 
	
    Given that Eq.~\eqref{eq:localderivative} ensures the group law~\eqref{eq:grouplaw}, it can be chosen as the basis for the perturbative RG treatment of a dynamical system. This same equation results from the application of the theory of envelopes
    ~\cite{kunihiro1995envelope, EFK, kunihirobook}. The left-hand side of Eq.~\eqref{eq:grouplaw} can be obtained from the perturbation theory for $t_1$ close to $t$, since the secular terms are small. Eq.~\eqref{eq:localderivative} then can improve the perturbative expansion, as described below.  
    
    To set up the perturbative RG for a nonlinear oscillator, we need the general form of its perturbative solution in the vicinity of $ t=t_1$. We show in Appendix~\ref{appendix:renormalizability} that a perturbative solution to Eq.~\eqref{eq:autonomousmodel} of order $O(\epsilon^{n_{max}})$ can always be written in the following form:
    \begin{widetext}
    \begin{align}\label{eq:generalsolution}
        y(t,t_1)=A e^{i \omega t}+\sum^{n_{max},m_{max}}_{n=1,m=0,m\neq1}\epsilon^n Y^{reg}_{nm}(A,A^*)e^{i m\omega t}+\sum^{n_{max},m_{max}}_{n=1,m=0}\epsilon^nY^{sec}_{nm}(A,A^*;t-t_1)e^{i m\omega t}+c.c.
    \end{align}
    In this solution, with two unknown $A,A^*$, we singled out non-secular terms oscillating at the prime frequency of the unperturbed ($\epsilon=0$) oscillator, $A e^{i \omega t}$, and $A^* e^{-i \omega t}$. Note that while the frequency is unperturbed, all orders of the perturbation theory can contribute to $A, A^*$, depending on how the initial conditions are implemented. The second term on the right hand side of Eq.~\eqref{eq:generalsolution} is a sum of non-secular terms of the perturbation theory with the corresponding oscillating exponentials. The summation index $n$ labels the order of the perturbation theory, and $m$ label the oscillation modes. The value of $m_{max}$ is determined by $n_{max}$, and a specific form of the perturbation, $f(y,\dot{y})$. The sum over $m$ possibly includes a constant term with $m=0$, but excludes the prime frequency, $m=1$. The prime frequency is not included since the non-secular terms oscillating at $\omega$ have already been taken into account in the first term on the right hand side, and its complex conjugate.  The third term on the right hand side is a sum of all secular terms to a given order. The prefactors of the oscillating exponentials are polynomials in $t$, which we chose to be functions of $t-t_1$, such that the secular terms vanish at $t_1=t$. 
    
    The crucial feature of solution~\eqref{eq:generalsolution} is that in the nonsecular terms the coefficients $Y^{reg}_{nm}(A,A^*)$ are independent of $t_1$ if the secular coefficients $Y^{sec}_{nm}(A,A^*; t-t_1)$ are polynomials of $(t-t_1)$ (rather than of $t$ and $t_1$ separately) of degree larger or equal to one. It will become apparent below that this fact ensures that the RG equations for the renormalized amplitude $A(t)$ do not contain time explicitly, and the expression for the renormalized solution does not contain secular terms. These two statements define renormalizability in the present context.  
    
    In order for $y(t,t_1)$ of Eq.~\eqref{eq:generalsolution} to coincide with $y(t,t_1;y(t_1),\dot{y}(t_1))$ in the vicinity of $t_1$, we need to impose the initial conditions on $y(t,t_1)$ as a function of its first argument: 
    \begin{align}\label{eq:initialconditions}
        &y(t_1,t_1)=y(t_1),\nonumber\\
        &\frac{dy(t,t_1)}{dt}|_{t\to t_1}=\dot{y}(t_1).
    \end{align}
    
    At this point we can assume that Eqs.~\eqref{eq:initialconditions} have been solved, and as a result two $t_1$-dependent amplitudes $A(t_1),A^*(t_1)$ were found. Importantly, this procedure never has to be carried out explicitly. If expressed through these amplitudes, $y(t_1)$ trivially satisfies 
    \begin{align}\label{eq:observable}
        &y(t_1)=A(t_1) e^{i \omega t_1}+\sum^{n_{max},m_{max}}_{n=1,m\neq1}\epsilon^n Y^{reg}_{nm}(A(t_1),A^*(t_1))e^{i m\omega t_1},
    \end{align}
    while the ``local" group law ~\eqref{eq:localderivative} implies that
    \begin{align}\label{eq:almostRG}
    \frac{d}{dt_1}\left[A(t_1) e^{i \omega t}+\sum^{n_{max},m_{max}}_{n=1,m=0,m\neq1}\epsilon^n Y^{reg}_{nm}(A(t_1),A^*(t_1))e^{i m\omega t}+\sum^{n_{max},m_{max}}_{n=1,m=0}\epsilon^nY^{sec}_{nm}(A(t_1),A^*(t_1);t-t_1)e^{i m\omega t}+c.c.\right]_{t_1\to t}=0.
    \end{align}
    Since Eq.~\eqref{eq:almostRG} must hold for all of $t$ after the limit $t_1\to t$ is taken, we can equate to zero derivatives with respect to $t_1$ of all the pre-exponential factors. \cb{In particular, for the prime frequency terms, $e^{\pm i\omega t}$, we obtain 
    \begin{align}\label{eq:intermediate1011}
        \frac{d}{dt_1}A(t_1)=-\frac{d}{dt_1}\left[\sum^{n_{max}}_{n=1}\epsilon^nY^{sec}_{n1}(A(t_1),A^*(t_1);t-t_1)\right]_{t_1\to t},
    \end{align}
    as by construction non-secular terms are absent at the prime frequency: $Y^{reg}_{n1}=0$. Only secular terms linear in $t-t_1$ will contribute to the right hand side of Eq.~\eqref{eq:intermediate1011} because of the $t_1\to t$ limit. }
    
    \cb{For notational convenience, we define 
    \begin{align}\label{eq:f(A)}
        F_n(A,A^*)=-\lim_{t_1\to t}\frac{d Y^{sec}_{n1}(A(t),A^*(t);t-t_1)}{dt_1},
    \end{align}
    which is a function of amplitudes $A,A^*$, but not their derivatives. Then we obtain the desired RG equation in the form of 
    \begin{align}\label{eq:dotA}
        \dot A(t)=\sum_{n=1}\epsilon^n F_n(A(t),A^*(t)).
    \end{align}}
    
    \cb{Equation.~\eqref{eq:dotA}, supplemented with Eq.~\eqref{eq:f(A)}, is the most general RG equations in the present context, while Eq.~\eqref{eq:observable} relates the renormalized solution of the differential equation~\eqref{eq:generalode} to the renormalized amplitudes $A(t), A^*(t)$. One can view Eq.~\eqref{eq:dotA} as a way to eliminate the secular terms from the perturbation theory by renormalizing the initial conditions~\cite{goldenfeld1996rg}. }
    
    \cb{Since Eq.~\eqref{eq:dotA} is obtained from the secular terms at the prime frequency, a comment on the fate of the secular terms at other frequencies is in order. Because of the hierarchical structure of the perturbation theory, the secular terms at non-prime frequencies appear only as a consequence of the secular terms at the prime frequency in lower orders of the perturbation theory, see Appendix~\ref{appendix:renormalizability} for details. This means that once Eq.~\eqref{eq:dotA} is enforced, and the secular terms at the prime frequency are eliminated, there is no need to consider secular terms at other frequencies, they are gone automatically. This can be explicitly seen below using the example of the Van der Pol oscillator: once one enforces Eq.~\eqref{eq:RGequationforVdP} for the renormalized amplitude to a given order of the perturbation theory, Eq.~\eqref{eq:almostRG} is automatically satisfied for the non-prime frequency terms, explicitly given in Eq.~\eqref{eq:PTforVdPfunctions}. }
    
    The RG equations derived above are one of the main results of this work. They are extremely simple to implement. At each order of the perturbation theory, while $A,A^*$ are still considered to be constants, one must choose the arbitrary coefficients of the general solution of the homogeneous equation in such a way as to ensure that the prime-frequency solution vanishes at $t=t_1$. As shown in the Appendix~\ref{appendix:renormalizability}, this guarantees the form of the perturbative solution stipulated in Eq.~\eqref{eq:generalsolution}, and RG equations~ \eqref{eq:dotA} and~\eqref{eq:f(A)} are trivially obtained. 
    
    \end{widetext}
    
    \subsection{Non-adiabatic RG equations}\label{sec:nonadiabaticRG}
    Below we will consider a situation in which the small parameters in a nonlinear oscillator problem are time-dependent, but this time dependence is slow, $\dot\epsilon/\omega\ll\epsilon$. Having in mind applications for calculation of geometric phases for oscillators, we will assume that these small parameters are taken through a cycle in the parameters space, and the duration of the cycle $\tau_c$ is such that $\dot\epsilon \tau_c\ll \epsilon$, even though $\omega\tau_c\gg1$. 
    
    For time-dependent $\epsilon(t)$, the RG equations will receive corrections, which can be expanded in powers of $\dot \epsilon$. In calculating these corrections, we restrict ourselves to the linear order in nonadiabaticity, since only such terms lead to the appearance of geometric phases. We neglect the small higher order derivatives of $\epsilon$ for the same reason. Under these conditions, one can obtain the nonadiabatic corrections in much the same way as the adiabatic ones were obtained. We expand the time-dependent small parameter as 
    \begin{align}\label{eq:epsilonexpansion}
        \epsilon(t)\approx \epsilon(t_1)+(t-t_1)\dot\epsilon(t_1), 
    \end{align}
    and treat $\dot\epsilon$ as a new time-independent small parameter. The resultant equation one has to solve is 
    	\begin{equation}\label{eq:nonautonomousmodel}
		\ddot{y} + \omega^2 y = [\epsilon(t_1)+(t-t_1)\dot\epsilon(t_1)] f(y,\dot{y}),
	\end{equation}
    in which $t_1$ should be viewed as a parameter. 
    
    It is clear from the preceding considerations, as well as from Appendix~\ref{appendix:renormalizability}, that the perturbative solution to Eq.~\eqref{eq:nonautonomousmodel} will have the same general form as prescribed by Eq.~\eqref{eq:generalsolution}, but $\epsilon^n$ replaced with $\epsilon^n\dot\epsilon^k$, with non-negative integers $n,k$ satisfying $n+k\geq 1$. 
    \begin{widetext}
     \begin{align}\label{eq:generalnonadiabaticsolution}
        y(t,t_1)=A e^{i \omega t}+\sum_{n+k\geq1,m=0,m\neq1}\epsilon^n\dot\epsilon^k Y^{reg}_{nkm}(A,A^*)e^{i m\omega t}+\sum_{n+k\geq1,m=0}\epsilon^n\dot\epsilon^k Y^{sec}_{nkm}(A,A^*;t-t_1)e^{i m\omega t}+c.c.
        \end{align}
    Functions $Y_{nkm}$ with $k=0$ in Eq.~\eqref{eq:generalnonadiabaticsolution} coincide with $Y_{nm}$ introduced in Eq.~\eqref{eq:generalsolution}. For the purpose of calculating the geometric phases, we only need the $k=0,1$ terms in Eq.~\eqref{eq:generalnonadiabaticsolution}. We will also neglect the $O(\epsilon^n\dot\epsilon)$ terms with $n>0$, even though these can be easily obtained, if needed. 
    
    The form of Eq.~\eqref{eq:nonautonomousmodel} shows that functions $Y_{01m}$ can be obtained from $Y_{10m}\equiv Y_{1m}$, which define the adiabatic RG equations, and the corresponding renormalized solution for $y(t)$ to $O(\epsilon)$ order. We obtain 
    \begin{align}
        &Y_{011}^{sec}=\left(\frac{i}{2\omega}+\frac12(t-t_1)\right)Y_{11}^{sec},\nonumber\\
        &Y_{01m\neq1}^{sec}=(t-t_1)Y^{reg}_{1m},\nonumber\\
        &Y_{01m\neq1}^{reg}=\frac{2im}{(m^2-1)\omega}Y_{1m}^{reg}.
    \end{align}
    Since only the linear in $t-t_1$ part of $Y_{011}^{sec}$ contributes to the RG equations, we obtain 
     \begin{align}\label{eq:fullRGequations}
        \dot A=\sum_{n=1}\epsilon^n F_n(A,A^*)+\frac{i\dot\epsilon}{2\omega} F_1( A, A^*).
    \end{align}
    In turn, the full renormalized solution, $y_R(t)$, of the original model~\eqref{eq:generalode} is 
    \begin{align}\label{eq:renormalizedobservable}
        &y_R(t)=A(t) e^{i \omega t_1}+\sum_{n=1,m\neq1}\epsilon^n Y^{reg}_{nm}(A(t),A^*(t))e^{i m\omega t_1}
        +\dot\epsilon\sum_{m\neq 1}\frac{2im}{m^2-1}Y^{reg}_{1m}(A(t),A^*(t))e^{im\omega t}+c.c.
    \end{align}
    \end{widetext}
    RG equations Eq.~\eqref{eq:fullRGequations}, and the equation for the renormalized solution~\eqref{eq:renormalizedobservable} for the nonlinear oscillator equation are one of the main results of this work. We emphasize that all of functions $Y^{sec}_{nm}$, $Y^{reg}_{nm}$, and the corresponding $F_n$ defined by Eq.~\eqref{eq:f(A)}, are obtained from simple perturbation theory. This makes the entire construction extremely easy to implement.

    We would like also to comment briefly on the case in which the frequency of the oscillator is time-dependent, providing only the corresponding correction to the RG equations. In this case a naive generalization of the approach we used for time-dependent $\epsilon$, Eq.~\eqref{eq:epsilonexpansion}, would not work, since $\omega(t_1)$ would make it to the oscillating exponentials in the perturbative solution, and enforcing the group law~\eqref{eq:localderivative} would lead to an explicit time-dependence in the obtained RG equations. 
    
    Instead, for a time-dependent $\omega(t)$, to obtain the adiabatic perturbative solution from Eq.~\eqref{eq:generalsolution} one must perform the following replacement in the phases of oscillating exponentials: 
    \begin{align}
        \omega t\to \int^t \d t'\, \omega(t'), 
    \end{align}
    while all the prefactors should be considered functions of the instantaneous value of $\omega(t)$. It is easy to show then that the equation for the $O(\epsilon^0\dot \omega)$ non-adiabatic correction at the prime frequency is given by 
    \begin{align}\label{eq:nonadiabaticequationomega}
        \ddot{y}_{na} + \omega^2 y_{na} = -i\dot\omega A e^{i\int^t\d t'\omega(t')}+c.c.
    \end{align}
    Using the same logic that led from Eq.~\eqref{eq:nonautonomousmodel} to the RG equation~\eqref{eq:fullRGequations}, we can write immediately that in the presence of a time-dependent frequency the RG equation becomes 
    \begin{align}\label{eq:fullRGequationswithomega}
        \dot A=F_\epsilon(A,A^*)+\frac{i\dot\epsilon}{2\omega} F_1(A,\dot A)-\frac{\dot\omega}{2\omega}A.
    \end{align}

    Note that the obtained $O(\epsilon^0\dot \omega)$ result is consistent with a well-known fact~\cite{LL1} that for a simple harmonic oscillator the ratio of its energy to its frequency is an adiabatic invariant. This implies that the combination $|A(t)|^2\omega(t)$ should not change with time for $\epsilon=0$, and thus $ F_\epsilon(A,A^*)=0$, which is consistent with Eq.~\eqref{eq:fullRGequationswithomega}.
    
    In specific examples, one may need to go beyond the $O(\epsilon^0\dot \epsilon)$ and $O(\epsilon^0\dot \omega)$ orders in the nonadiabatic RG equations. We will encounter a case where one needs to iterate beyond $O(\epsilon^0\dot \omega)$ in Section~\ref{sec:VdP} for the Van der Pol oscillator with a time-dependent frequency. We will illustrate the procedure for that specific case. 
    
    \section{Geometric phase for nonlinear oscillators}\label{sec:phase}
    
     In this Section, 
	we show how to use the formalism developed in Section~\ref{sec:RG} to describe the geometric phase for a nonlinear oscillator with a limit cycle. In Section ~\ref{sec:VdP} below we will apply this formalism to Van der Pol-type oscillators. 
	
	Upon using the polar form $A=r e^{i\theta}/2$ (note the factor of 1/2), Eq.~\eqref{eq:fullRGequations} can be used to write equations for is magnitude and phase, whose general form is
	\begin{equation}\label{eq:dotrdottheta}
	\frac{\d r}{\d t} = f(r, \bm{\epsilon}, \dot{\bm{\epsilon}}), \quad \frac{\d \theta}{\d t} = \Omega(r,\bm{\epsilon},\dot{\bm{\epsilon}})
	\end{equation}
	\cb{Here $\bm{\epsilon} = \{\epsilon_1, \epsilon_2\}$ should be thought of as a vector of small parameters, which change adiabatically through a cycle in the parameter space, but we will keep a single name $\bm{\epsilon}$ for these parameters. }
 
 In coordinates $(r,\theta)$, the equation for a limit cycle can be written as $f(R(\bm{\epsilon}), \bm{\epsilon}, 0) = 0$, where $R(\bm{\epsilon})$ is the radius of the cycle. If the parameters are changing adiabatically, the deviations from the cycle will be small. Following the approach taken in \cite{kepler1991geometric}, we introduce the variable $z = r - R(\bm{\epsilon})$. Then up to the first order in $z$ and $\dot{\bm{\epsilon}}$ we get:
	\begin{equation}
		\frac{\d z}{\d t} = z f'_r(R, \bm{\epsilon}, 0) + \dot{\bm{\epsilon}}\cdot f'_{\dot{\bm{\epsilon}}}(R,\epsilon,0) - \dot{\bm{\epsilon}}\cdot R'_{\bm{\epsilon}},
		\label{eqz}
	\end{equation}
	\begin{equation}
	 \frac{\d \theta}{\d t} = \Omega(\bm{\epsilon}, 0) + z \Omega'_r(\bm{\epsilon}, 0) + \dot{\bm{\epsilon}}\cdot\Omega'_{\dot{\bm{\epsilon}}}(\bm{\epsilon}, 0).
  \label{eqtheta}
	\end{equation}
	It can be shown \cite{kepler1991geometric} that there is a vector function $\bm{\xi}(\bm{\epsilon}, \dot{\bm{\epsilon}})$ such that up to the first order $z$ can be written as ${z = \dot{\bm\epsilon}\cdot\bm{\xi}(\bm{\epsilon}, \dot{\bm{\epsilon}})}$. Therefore, we see that $\dot{z} = 0$ up to the first order in $\dot{\bm{\epsilon}}$. Substituting it into Eq.~(\ref{eqz}), we get an equation for $\bm{\xi}$:
	\begin{equation}
		0 = \bm{\xi} f'_r(R, \bm{\epsilon}, 0) + f'_{\dot{\bm\epsilon}}(R,\bm{\epsilon},0) - R'_{\bm{\epsilon}}.
	\end{equation}
	Substituting $\bm{\xi}$ in Eq. (\ref{eqtheta}) we get:
	\begin{equation}\label{eq:thetadot}
		\frac{\d \theta}{\d t} = \Omega(\bm{\epsilon}, 0) + \dot{\bm\epsilon}\cdot \frac{R'_{\bm\epsilon} - f'_{\dot{\bm\epsilon}}(R, \bm{\epsilon}, 0)}{f'_r(R, \bm{\epsilon}, 0)} \Omega'_r(\bm{\epsilon}, 0) + \dot{\bm{\epsilon}}\cdot \Omega'_{\dot{\bm{\epsilon}}}(\bm{\epsilon}, 0).
	\end{equation}   
	The first term in the right hand side of Eq.~\eqref{eq:thetadot} describes the accumulation of the dynamical part of the phase, $\theta_{dyn}$, such that 
	\begin{align}\label{eq:dynamicphase}
	    \theta_{dyn}=\int^t dt'\, \Omega(\bm{\epsilon}(t'),0). 
	\end{align}
	The dynamical phase in general scales linearly with the time it takes the system to complete a cycle in the parameter space. In turn, 
	last two terms correspond to geometric phase:
	\begin{equation}\label{eq:thetageom}
		\theta_{geom} = \int\left(\frac{R'_{\bm\epsilon} - f'_{\dot{\bm\epsilon}}(R, \bm{\epsilon}, 0)}{f'_r(R, \bm{\epsilon}, 0)} \Omega'_r(\bm{\epsilon}, 0)+\Omega'_{\dot{\bm\epsilon}}(\bm{\epsilon}, 0)\right)\cdot\d\bm{\epsilon}.
	\end{equation}
	This contribution to the phase does not depend on the total time of travel along a closed path in the parameter space, as long as this time is not too short. \cb{We can write
	\begin{align}\label{eq:thetageom2}
	    \theta_{geom}= \int\vartheta\equiv\int (a_1(\bm{\epsilon})\d\epsilon_1+a_2(\bm{\epsilon})\d\epsilon_2),
	\end{align}
	where $a_{1,2}(\bm{\epsilon})$ are read off from Eq.~\eqref{eq:thetageom}. }
	
	Of physical interest is the accumulation of phase along a closed curve, $\gamma$, in the parameter space, in which case the curve is a boundary of some region $G$: $\gamma = \partial G$.  \cb{To compute the geometric phase, we utilize the Stokes' theorem in Eq.~\eqref{eq:thetageom2}:
	\begin{equation}
		\theta_{geom} = \oint\limits_{\gamma} \vartheta = \int\limits_{G} \chi\, \d\epsilon_1\wedge\d\epsilon_2,
	\end{equation} 
	where 
	\begin{equation}
		\chi = \frac{\partial a_2}{\partial \epsilon_1} - \frac{\partial{ a_1}}{\partial \epsilon_2}
	\end{equation}
	is the analog of the ``Berry curvature'' in the present problem~\cite{berry1984phase}.} Calculation of this curvature for Van der Pol-type oscillators is the primary goal of the following Section.
	
	\section{Van der Pol and Van der Pol-Duffing oscillators}\label{sec:VdPoscillators}
	To illustrate the results obtained in Sections~\ref{sec:RG} and~\ref{sec:phase}, we consider the Van der Pol-Duffing (VdPD) oscillator, defined via a differential equation for its displacement, $y(t)$:
	\begin{equation}
		\ddot{y} + \omega^2 y = \mu (1-y^2)\dot{y} - \beta y^3.
		\label{eq:VdPD}
	\end{equation}
	In this equation, the term with $\omega^2$ describes the linear force acting on the oscillator, the one proportional to $\beta$ is a nonlinear correction to the force, and the coefficient proportional to $\mu$ is a nonlinear dissipative-like term. This term provides dissipation for $|y|>1$, but leads to generation for $|y|<1$. It is well known that the nonlinear dissipative term leads to the existence of a limit cycle in this model, which roughly corresponds to the trajectory in the phase space for which the dissipation and generation exactly balance each other out.  The Van der Pol (VdP) oscillator model is obtained from the VdPD one by setting $\beta\to 0$. It is worth keeping in mind that since $y$ is chosen to be dimensionless in Eq.~\eqref{eq:VdPD}, $\mu$ and $\beta^{1/2}$ have the same units as the frequency $\omega$. In this work we are interested in the non-autonomous version of VdP and VdPD oscillators, in which parameters $\omega$, $\mu$, and $\beta$ have time dependence, albeit very slow one. 
	
	Recently, a way to study the geometric phases in dissipative systems was proposed by their ``Hamiltonization" in Ref.~\cite{chakraborty2018hannay}. In this work, we adopt the perturbative renormalization group to study changes in the limit cycle of the VdPD oscillator under slow time evolution of its parameters, and to determine the associated geometric phase.  We will start with the VdP model to reproduce some known results using the RG. Later we will turn to the VdPD model, and show that in that case the curvature in the parameter space is singular and scales as $1/\mu^2$.

	\subsection{Perturbative RG for the VdP model}\label{sec:VdP}
	
	While developing a perturbative expansion in $\mu$ for the solution of Eq.~\eqref{eq:VdPD} with $\beta=0$ (VdP model), we always work up to the lowest order to which the desired phenomenon - geometric phases in our case - appear. It turns out that for the VdP model going to $O(\mu^2)$ order is sufficient, as will be discussed below.  The corresponding expressions for the perturbation theory solution are most easily obtained using some software for symbolic computation, of which we used Wolfram \textit{Mathematica}.

	\subsubsection{Adiabatic RG equations}
	It is apparent from Eqs.~\eqref{eq:dotA} and~\eqref{eq:f(A)} that the adiabatic RG equations are determined by the linear in $(t-t_1)$ secular terms at the prime frequency. To obtain these terms, we organize the perturbation theory as 
 \begin{align}\label{eq:PTforVdP}
    y(t,t_1)=y_0(t)+\mu\, y_1(t,t_1)+\mu^2 y_2(t,t_1), 
\end{align}
where 
 \begin{widetext}
\begin{align}\label{eq:PTforVdPfunctions}
    y_0={}&A e^{i \omega t}+A^* e^{-i \omega t},\nonumber\\
    y_1={}&\underline{\frac{1}{2} (t-t_1) A \left(1-|A|^2\right)  e^{i \omega t}}+\frac{iA^3}{8 \omega }e^{3 i  \omega t }+c.c.,\nonumber\\
    \begin{split}
    y_2={}& \left(\underline{-\frac{i}{16\omega}(t-t_1) A(2 - 8|A|^2+7|A|^4)} + \frac{1}{8} (t-t_1)^2 A(1-4|A|^2+3|A|^4)\right)e^{i\omega t} + \\&
    + \left(-\frac{1}{64\omega^2} A^3(2+|A|^2)+\frac{3i}{16\omega}(t-t_1)A^3(1-|A|^2)\right)e^{3i\omega t} - \frac{5}{192\omega^2} A^5 e^{5i\omega t} + c.c.
    \end{split}
\end{align}
 \end{widetext}
 The linear in $t-t_1$ secular terms at the prime frequency in each order of the perturbation theory are underlined.
 
 The equation for the renormalized amplitude $A(t)$ can be obtained from Eqs.~\eqref{eq:PTforVdPfunctions}, using Eqs.~\eqref{eq:dotA} and~\eqref{eq:f(A)}: 
	\begin{align}\label{eq:RGequationforVdP}
	    \dot{A}=\frac{\mu}{2} A \left(1-|A|^2\right)-\frac{i\mu^2}{16 \omega} A \left(2-8|A|^2+7 |A|^4\right).
	\end{align}
	Upon using the polar representation $A=r e^{i\theta}/2$ (note the factor of $1/2$), the equations for $r$ and $\theta$ become
	\begin{align}
	    &\dot r=\frac{\mu}{8}r(4-r^2),\nonumber\\
	    &\dot \theta=-\frac{ \mu ^2}{256 \omega } \left(7 r^4-32 r^2+32\right).
	\end{align}
	These equations lead to the well-known results~\cite{nayfehbook}: the VdP model has a limit cycle of radius $r=2$, defined as $r$ for which $\dot r=0$, and at the limit cycle the correction to the frequency, given by $\dot\theta$ evaluated for $r=2$, is $-\mu^2/16\omega$. 
	
	It is instructive to consider the RG equations in the vicinity of the limit cycle, since it provides physical picture behind the needed nonadiabatic corrections to the equations. For $r=2+\delta r$, $\delta r\ll1$, we obtain 
	\begin{align}\label{eq:RGnearcycle}
	    &{\delta \dot r}=-\mu\delta r,\nonumber\\
	    &\dot \theta=-\frac{ \mu ^2}{16 \omega } -\frac{3 \mu ^2 }{8 \omega } \delta r.
	\end{align}
	Equations~\eqref{eq:RGnearcycle} show that $\mu$ plays the role of the relaxation rate back onto the limit cycle, and $1/\mu$ is the corresponding relaxation time. Therefore, an $O(\mu^0\dot\omega)$ nonadiabatic correction to the equation for $\dot r$ will produce a deviation from the limit cycle amplitude $\delta r\propto \dot\omega/\mu$. In turn, this will lead to $O(\mu,\dot\omega)$ correction to the frequency of the oscillation, linear both in $\mu$ and $\dot\omega$. This means that one needs to also take into account $O(\mu\dot\omega)$ nonadiabatic corrections to $\dot\theta$ to have a consistent treatment. These corrections are discussed below.  
	
	\subsubsection{Nonadiabatic RG equations}
	The leading nonadiabatic corrections to $O(\mu^0\dot\mu)$ and $O(\mu^0\dot\omega)$ orders can be simply read off Eq.~\eqref{eq:fullRGequationswithomega}. 
	\begin{align}
	    &\dot r=\frac{\mu}{8}r(4-r^2)-\frac{\dot\omega}{2\omega}r,\nonumber\\
	    &\dot \theta=-\frac{ \mu ^2}{256 \omega } \left(7 r^4-32 r^2+32\right)+\frac{\dot\mu}{16\omega}(4-r^2) .
	\end{align}
	As discussed above, the equation for $\dot \theta$ needs to be corrected with $O(\mu\dot\omega)$ terms. For the present case, these can be easily obtained by noting that the solution to the VdP model obtained from amplitude $A$ with $O(\mu^0\dot\omega)$ corrections by construction solves the original equation to up to $O(\mu^2)$ and $O(\mu^0\dot\omega)$ order. We can iterate RG equations to order $O(\mu\dot\omega)$ by substituting the solution up to $O(\mu\dot\omega)$ order into the nonlinear term, and collecting the $O(\mu\dot\omega)$ secular terms, and only the part of those that makes a contribution to the $\dot\theta$ equation. The result is 
	\begin{align}
	    &\dot r=\frac{\mu}{8}r(4-r^2)-\frac{\dot\omega}{2\omega}r,\nonumber\\
	    &\dot \theta=-\frac{ \mu ^2}{256 \omega } \left(7 r^4-32 r^2+32\right)
	    +\frac{4-r^2}{16\omega}\dot\mu-\frac{\mu}{16\omega^2}r^2\dot\omega.
	\end{align}
	At this point we can apply the general equation~\eqref{eq:thetageom} to $\bm{\epsilon}=(\mu,\omega)$ to obtain near the limit cycle, $r=2$:
 \be
 \theta_{geom}=\int(a_\mu \d\mu+a_\omega\d \omega),
 \ee
 with
	\begin{align}
	    a_\mu=0,\quad a_\omega=\frac{\mu}{8\omega^2}.
	\end{align}
 The corresponding curvature is 
 \begin{align}
 \chi_{{\rm vdp}}={\partial_\mu a_\omega}-\partial_\omega a_\mu=\frac{1}{8\omega^2}.
 \label{curvatureVdP}
 \end{align}
 The above results are different in sign as compared with those obtained in  Ref.~\cite{chakraborty2018hannay} using ``Hamiltonianization" of the VdP model. We have confirmed our results via numerical simulations. To this end, we numerically solved Eq.~\eqref{eq:VdPD} with $\beta=0$ for a long enough time, including a cycle of changing $\mu$ and $\omega$ along a circle in the parameter space. During the cycle, the two parameters were changed according to 
 \begin{align}
    \mu(t)&=\mu_0+\delta\mu\left[\cos\left(2\pi \frac{t}{T}\right)-1\right],\nonumber\\
    \omega(t)&=\omega_0+\delta\omega\sin\left(2\pi \frac{t}{T}\right),
    \label{counterclockwise}
 \end{align}
 with $\delta\mu\ll\mu_0$, $\delta\omega, T^{-1}\ll\omega_0$. Note that $\mu(0) = \mu_0$, so both parameters change continuously. Under such circumstances, for large enough $T$, one expects the geometric phase to be given by 
 \begin{align}\label{eq:theoreticalphase}
     \theta_{\rm{geom}}\approx \pi\delta\omega\delta\mu\chi_{{\rm VdP}}(\mu_0,\omega_0). 
 \end{align}

\begin{figure}[t]
    \centering
    \includegraphics{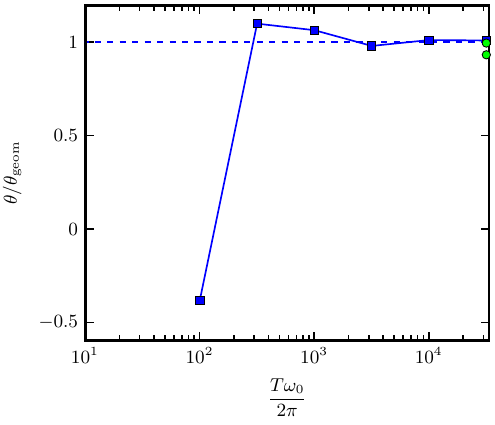}
    \caption{Phase accumulated during a cycle in the $(\mu,\omega)$ space as a function of the time to complete the cycle, $T$ (blue squares). The parameters used for the simulation are $\omega_0=2$, $\delta \omega = 0.2$, $\mu=0.1$, $\delta \mu = 0.01$. Geometric phase calculated using Eq.~\eqref{eq:theoreticalphase} is $\theta_{\rm{geom}} = 1.98\cdot10^{-4}$. Green circles: the geometric phase obtained for the longest cycle, $\omega_0 T=2\cdot10^5$, and different values of $\mu=0.2, 0.05$, and unchanged values of $\delta\mu, \delta\omega, \omega_0$.}
    \label{fig:VdP}
\end{figure}

  The evolution time before the cycle must be long enough for the limit cycle to get established, and long enough as compared with the relaxation time toward the limit cycle, $1/\mu$. To extract the geometric part of the phase, we consider evolution for the same cycle traversed in the two opposite directions, since the difference between the corresponding phases will not contains the dynamic part, Eq.~\eqref{eq:dynamicphase}. Eq.~\eqref{counterclockwise} corresponds to a path traversed in the counterclockwise (positive) direction. Since the change in the phase in both cases is very small, we pick a particular period of oscillation after the cycle is completed, and determine the time stamp when $y(t) = 0$. Given the time stamps $t_+$ and $t_-$ that correspond to counterclockwise and clockwise directions, we can calculate the geometric part of the phase as 
  \begin{align}\label{eq:phasedifference}
      \theta= \frac{\omega_0(t_- - t_+)}{2}.
  \end{align}
  The numerical results are presented in Fig.~\ref{fig:VdP}, which shows that the phase defined in Eq.~\eqref{eq:phasedifference} saturates at the theoretical value for the geometric phase~\eqref{eq:theoreticalphase} (in both sign and magnitude) for long enough cycle in the parameter space.  

\subsection{Perturbative RG for the VdPD model}

We now turn our attention to the Van der Pol-Duffing model. We consider Eq.~\eqref{eq:VdPD} with $\omega(t)={\rm const}$, and $\beta\neq 0$. As will become clear shortly, there is a singular in $\mu$ contribution to the Berry curvature that scales as $\chi_{\rm vdp}\propto\beta/\mu^2$. We will limit ourselves to obtaining only such singular contribution solely for the sake of clarity of presentation. Obtaining other terms in the curvature presents no difficulty using the present approach. We note that keeping only the singular term in the Berry curvature is consistent if $\beta/\mu^2\gg1$, while the condition that the system does not deviate from the limit cycle too much is given by $\beta/\omega^2\ll1$. Therefore, the results obtained below are valid for $\mu^2\ll \beta\ll\omega^2$.

The needed perturbation series takes the form 
\begin{align}\label{eq:PTforVdPD}
    y(t,t_1)=y_0(t)+\mu y_{10}(t,t_1)+\beta y_{01}(t,t_1) + \mu\beta y_{11}(t,t_1), 
\end{align}
where
\begin{widetext}
\begin{align}
    y_0={}&A e^{i \omega t}+A^* e^{-i \omega t},\nonumber\\
    y_{10}={}&\underline{\frac{1}{2} (t-t_1) A \left(1-|A|^2\right)}  e^{i\omega t}+\frac{i A^3  }{8\omega }e^{3 i  \omega t }+c.c.,\nonumber\\
    y_{01}={}&\underline{\frac{3i}{2\omega} (t-t_1) A |A|^2 }  e^{i\omega t}+\frac{A^3}{8 \omega^2}e^{3i\omega t}+c.c.,\nonumber\\
    \begin{split}
    y_{11}={}& \left(\underline{-\frac{1}{4\omega^2} (t-t_1)A|A|^2(3-2|A|^2)} + \frac{3i}{2\omega}(t-t_1)^2A|A|^2(1-|A|^2)\right)e^{i\omega t} + \\&
    + \left(\frac{3i}{32\omega^3}A^3(1-2|A|^2) + \frac{3}{16\omega^2}(t-t_1)A^3(1-4|A|^2)\right)e^{3i\omega t} + \frac{i}{24\omega^3}A^5 e^{5i\omega t}+ c.c.
    \end{split}
\end{align}
\end{widetext}
The secular terms at the prime frequency are again underlined. 

\begin{figure}[t]
    \centering
    \includegraphics{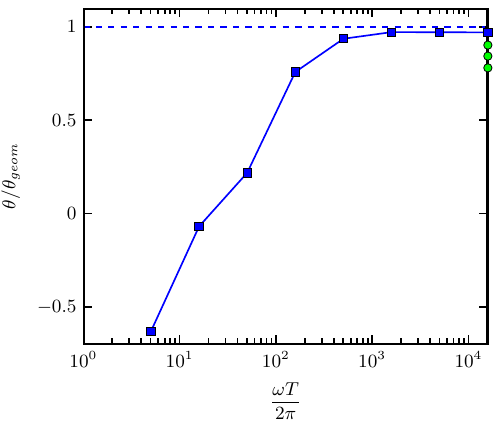}
    \caption{Geometric phase accumulated during a cycle in the $(\mu,\beta)$ space as a function of the time to complete the cycle, $T$ (blue squares). The parameters used for the simulation, in appropriate units, are $\beta=0.005$, $\delta \beta = 0.001$, $\mu=0.01$, $\delta \mu = 0.0005$, $\omega = 1$. Geometric phase calculated using Eq. \eqref{curvatureVdPD}: $\theta_{geom} = 1.31\cdot10^{-4}$. Green circles: the geometric phase obtained for the longest cycle, $\omega T= 10^{5}$, and different values of $(\mu,\beta)=(0.015, 0.01125),(0.02, 0.02),(0.025, 0.03125)$, for which $\beta/\mu^2=50$.}
    \label{fig:VdPD}
\end{figure}

We can read off both the adiabatic and nonadiabatic RG equations from Eqs.~\eqref{eq:fullRGequations} and~\eqref{eq:f(A)}:
		\begin{align}
	    \dot{A}&=\frac{\mu}{2} A \left(1-|A|^2\right)+\frac{3i\beta}{2\omega} A |A|^2-\frac{\mu\beta}{4\omega^2}A|A|^2(3-2|A|^2) \nonumber\\
	    &+\frac{i\dot\mu}{4\omega}A\left(1-|A|^2\right)-\frac{3\dot\beta}{4\omega^2}A|A|^2.
	\end{align}
 Switching to the polar representation, $A=r e^{i\theta}/2$, we obtain the following equations for $\dot r$ and $\dot\theta$:
		\begin{align}
	    \dot r &= \frac{\mu}{8}  r\left(4-r^2\right) - \frac{\mu\beta}{32\omega^2}r^3(6-r^2)- \frac{3\dot{\beta}}{16\omega^2}r^3\nonumber\\
	    \dot\theta &= \frac{3\beta}{8\omega}r^2 + \frac{\dot{\mu}}{16\omega}(4-r^2).
	\end{align}
These equations have the form described in Eq.~\eqref{eq:dotrdottheta}, so we can use the results of Section~\ref{sec:phase} for $\bm{\epsilon} = (\mu, \beta)$. In general, 
% From Eq (\ref{Rrhs}), (\ref{thetarhs}) we see, that $f'_{\dot{\mu}} = 0, \Omega'_{\dot{\beta}} = 0$, therefore we get:
	\begin{equation}
		\theta_{\rm{geom}} =\int(a_\mu\d\mu+a_\beta \d\beta).
	\end{equation} 	
As we mentioned before, we retain only the terms in the connection and curvature that are singular in the limit $\mu\to 0$, so in this sense 
	\begin{equation}
		a_\mu \approx 0, \quad a_\beta \approx -\frac{3\beta}{2\mu\omega^3}.
		\label{connections}
	\end{equation}
For the curvature we obtain
	\begin{equation}
		\chi_{{\rm vdpd}} = \partial_\mu a_\beta-\partial_\beta a_\mu=\frac32 \frac{\beta}{\mu^2\omega^3}.
  \label{curvatureVdPD}
	\end{equation}

We confirmed the result for the curvature in the parameter space in the case of the Van der Pol-Duffing model via numerical simulations identical to those for the Van der Pol model, but with $\beta$ replacing $\omega$ as one of the parameters whose change drives the accumulation of the geometric phase. That is, the cycle in the parameter space was parametrized according to 
\begin{align}
    \mu(t)&=\mu_0+\delta\mu\left[\cos\left(2\pi \frac{t}{T}\right)-1\right],\nonumber\\
    \beta(t)&=\omega_0+\delta\beta\sin\left(2\pi \frac{t}{T}\right),
    \label{counterclockwise}
 \end{align}
Numerical results are presented in Fig. \ref{fig:VdPD}.

 \section{Discussion}\label{sec:discussion}
 In this paper we presented a formulation of the renormalization group treatment of nonlinear oscillators. It allows to improve the perturbation theory for the oscillator, which contains secular terms.  Our approach is based on the exact group law~\eqref{eq:grouplaw} satisfied by the solution of the differential equation~\eqref{eq:autonomousmodel} describing a nonlinear oscillator. 

The way of deriving the renormalization group equations from a (singular) perturbative solution of a nonlinear oscillator problem is essentially algorithmic, and is summarized by Eqs.~\eqref{eq:fullRGequationswithomega}, and~\eqref{eq:f(A)}. We demonstrated that the nonlinear oscillator models are perturbatively renormalizable in Appendix~\ref{appendix:renormalizability}. 

We used the obtained RG equations to consider the appearance of geometric phase in dissipative oscillator models with limiting cycles. Specific examples of application of the developed formalism included the Van der Pol, and Van der Pol-Duffing oscillators. The results for the geometric phase obtained using the renormalization group treatment were in good agreement with numerics.

\cb{The present treatment of the nonlinear oscillator problem is inspired by previous works, especially Refs.~\cite{goldenfeld1996rg,kunihiro1995envelope}. It shows that the perturbative RG for nonlinear oscillators can be a practical tool to calculate geometric phases in dynamical systems with limit cycles. Our treatment can be shown to work perfectly well for the models mentioned in Ref.~\cite{Oono} as those for which the approach of Ref.~\cite{goldenfeld1996rg} requires refinements. }

 \section{Acknowledgements}
We are grateful to Eleftherios Kirkinis for useful discussions, and in particular for bringing Ref.~\cite{goto1999lie} to our attention as this paper was being prepared for publication. The work of DAP was supported by the National Science Foundation Grant No. DMR-2138008.

\appendix

\section{Renormalizability of the nonlinear oscillator}\label{appendix:renormalizability}

In this Appendix we show that a perturbative solution to model~\eqref{eq:autonomousmodel} can be written in the form given by Eq.~\eqref{eq:generalsolution}, which shows renormalizability of the nonlinear oscillator model. We also present a simple constructive way to obtain that form of the perturbative solution in practical calculations. In particular, that form naturally arises when one obtains the solution via some symbolic calculation software, like Wolfram \textit{Mathematica}.  
\begin{widetext}

For clarity, we reproduce Eq.~\eqref{eq:generalsolution} here: 
\begin{align}\label{eq:generalsolutionAppendix}
        y(t,t_1)=A e^{i \omega t}+\sum^{n_{max},m_{max}}_{n=1,m=0,m\neq1}\epsilon^n Y^{reg}_{nm}(A,A^*)e^{i m\omega t}+\sum^{n_{max},m_{max}}_{n=1,m=0}\epsilon^nY^{sec}_{nm}(A,A^*;t-t_1)e^{i m\omega t}+c.c.
    \end{align}
We now solve the original problem with time-independent parameters order by order in $\epsilon$:
\begin{equation}\label{eq:appendixgeneralode}
		\ddot{y} + \omega^2 y = \epsilon f(y,\dot{y}).
	\end{equation}
In doing so, we have in mind a pertubative expansion in the vicinity of $t=t_1$.
The unperturbed solution is 
\begin{align}\label{eq:appendixy0}
    y_0=Ae^{i\omega t}+A^*e^{-i\omega t}. 
\end{align}
The two unknown amplitudes $A,A^*$ are sufficient to match any Cauchy data, so we are free to eliminate all arbitrary constants appearing in higher-order terms to our liking. 

The first order correction is obtained by substituting $y_0$ into the nonlinearity in the right hand side of Eq.~\eqref{eq:appendixgeneralode}. The function $f(y,\dot y)$ is assumed to be analytic in both of its arguments, and to admit a Taylor expansion in them around $y=0,\,\dot y=0$. Thus we will think about it as a polynomial of $y$ and $\dot y$. To linear order in $\epsilon$, we can write $y=y_0+\epsilon y_1$, with an $\epsilon$-independent $y_1$, which satisfies the following equation: 
\begin{align}\label{eq:appendixfirstorderode}
    \ddot{y}_1 + \omega^2 y_1 =\sum_{m=0}f_m(A,A^*)e^{im\omega t}+c.c.
\end{align}
The specific form of the coefficients $f_m(A,A^*)$, of course, depends on the specific nonlinearity in the problem. The solution of inhomogeneous equation~\eqref{eq:appendixfirstorderode} is a sum of the general solution of the homogeneous equation, and a particular solution of the inhomogeneous equation. The former has the same form as~\eqref{eq:appendixy0}, but with some new constants $C_1$ and $C_1^*$, while the inhomogeneous solution  contains secular terms at the prime frequency, as well as regular terms at all other frequencies: 
\begin{align}\label{eq:appendixy1}
    y_1(t)=C_1e^{i\omega t}-\frac{i}{2\omega}f_1 te^{i\omega t}+\sum_{m\neq1}\frac{ f_m}{(1-m^2)\omega^2}e^{im\omega t}+c.c.
\end{align}
In the above expression we explicitly separated the secular term at the prime frequency. We can now choose the arbitrary constant $C_1$ in such a way as to make the secular term vanish at $t=t_1$, which amounts to setting
\begin{align}\label{eq:appendixC1}
    C_1=\frac{i}{2\omega}f_1(A,A^*)t_1. 
\end{align}
This brings the first order of the perturbation theory to the form specified by Eq.~\eqref{eq:generalsolutionAppendix}. We show below that iteration of the above procedure for the first order perturbation theory is guaranteed to bring the entire perturbative series to the form of Eq.~\eqref{eq:generalsolutionAppendix}.

We repeat the procedure outlined for the first order perturbation theory to any order in $\epsilon$. Since 
\begin{align}
    \frac{d}{dt}\left((t-t_1)^le^{im\omega t}\right)=l(t-t_1)^{l-1}e^{im\omega t}+im\omega(t-t_1)^le^{im\omega t}
\end{align}
contains prefactors in front of the oscillating exponentials that are functions of $(t-t_1)$ only, the general term in the inhomogeneous solution appearing in higher orders of the perturbation theory (labeled with $n$) will be obtained from 
\end{widetext}
\begin{align}\label{eq:appendixgeneralinhomogeneity}
    \ddot y_n+\omega^2 y_n=B_{ml}(A,A^*)(t-t_1)^le^{i m \omega t},
\end{align}
in which both $l$ and $m$ are non-negative integers. We emphasize that the right hand side of Eq.~\eqref{eq:appendixgeneralinhomogeneity} is meant to represent a typical inhomogeneity appearing to order $\epsilon^n$. In general the inhomogeneity will contain several terms with different $m,l$, all of which can be treated in the way described below. 

For $m\neq 1$, we can seek a solution to Eq.~\eqref{eq:appendixgeneralinhomogeneity} in the form of $y_n(t)=P_{ml}(t-t_1)e^{i m\omega t}$, where $P_{ml}(t-t_1)$ is a polynomial of $(t-t_1)$, satisfying 
\begin{align}\label{eq:appendixpolynomialmneq1}
\ddot{P}_{ml}+2im\omega\dot{P}_{ml}+(1-m^2)\omega^2P_{ml}=B_{ml}(t-t_1)^l.
\end{align}
Because of a non-vanishing term without a time derivative in the left hand side of Eq.~\eqref{eq:appendixpolynomialmneq1}, it is clear that $P_{ms}(t-t_1)$ is a polynomial of degree $l$:
\begin{align}
P_{ml}(t-t_1)=\sum_{k=0}^{l}p_{k}(t-t_1)^k.
\end{align}

Then \begin{align}
p_l=\frac{B_{ml}}{(1-m^2)\omega^2},\,\,\, p_{l-1}=-\frac{2imlB_{ml}}{(1-m^2)^2\omega^3},
\end{align}
and for $k\leq l-2$ we obtain a recursive relation
\begin{align}\label{eq:appendixrecursionmneq1}
(k+2)(k+1)p_{k+2}+2im\omega(k+1)p_{k+1}+(1-m^2)\omega^2p_k=0.
\end{align}
The recursive relation terminates at $k=0$, determining uniquely all the coefficients. We thus conclude that the non-prime frequencies terms in the perturbation series all have prefactors in front of the oscillating exponentials that are polynomials of $(t-t_1)$ only (as opposed to depending separately on $t_1$, $t$), and conform to the form prescribed by Eq.~\eqref{eq:generalsolutionAppendix}.

Turning our attention to $m=1$, we note that if we seek a particular solution in the form of $y_n(t)=P_{1l}(t-t_1)e^{i \omega t}$, then $P_{1l}$ satisfies 
\begin{align}
\ddot{P}_{1l}+2i\omega\dot{P}_{1l}=B_{1l}(t-t_1)^l,
\end{align}
and thus $P_{1l}$ is a polynomial of degree $l+1$:
\begin{align}
P_{1l}(t-t_1)=\sum_{k=0}^{l+1}p_{k}(t-t_1)^k.
\end{align}
In this case 
\begin{align}
    p_{l+1}=-\frac{iB_{1l}}{2(l+1)\omega},
\end{align} 
and for $k\leq l$ we have a recursion relation
\begin{align}\label{eq:appendixrecursion1}
(k+1)kp_{k+1}+2i\omega k p_{k}=0.
\end{align}
This recursion relation terminates at $k=1$, leaving $p_0$ undetermined. The coefficient $p_0$ can always be set to zero by choosing an appropriate solution of the homogeneous equation, the way it was done in Eqs.~\eqref{eq:appendixy1} and~\eqref{eq:appendixC1}. Therefore, we conclude that the prime-frequency terms in the perturbative solution, apart from the solution to the homogeneous equation, Eq.~\eqref{eq:appendixy0}, have prefactors in front of the oscillating exponentials that are polynomials of $(t-t_1)$, which also vanish at $t=t_1$, such that there are no regular terms, only secular, in the perturbative solution at the prime frequency. This proves that  ~\eqref{eq:generalsolutionAppendix} or~\eqref{eq:generalsolution} represent a general perturbative solution to the nonlinear oscillator model, and also provides a constructive way to obtain this solution by choosing the coefficients of the solution of the homogeneous equation at every order of the perturbation theory in such a way that there are no regular terms, only secular, at the prime frequency. 

We also would like to point out that recursion relations~\eqref{eq:appendixrecursionmneq1} and~\eqref{eq:appendixrecursion1} show that all the higher-order secular terms are fully determined by the linear in $(t-t_1)$ ones appearing at the prime frequency in each order of the perturbation theory. It is no wonder than only these linear in $(t-t_1)$ secular terms at the prime frequency fully determine the RG equation~\eqref{eq:fullRGequations}.

\bibliography{rgvdp}
\bibliographystyle{apsrev4-2}

\end{document}